\documentstyle[12pt,fleqn,epsfig]{article}
\begin{document}
\begin{center}
{\bf INSTITUT~F\"{U}R~KERNPHYSIK,~UNIVERSIT\"{A}T~FRANKFURT}\\
D - 60486 Frankfurt, August--Euler--Strasse 6, Germany
\end{center}

\hfill IKF--HENPG/7--97

\vspace{1cm}

\begin{center}

{\Large \bf A Method to Study "Chemical" Fluctuations }
\vspace{0.3cm}
{\Large \bf in  Nucleus--Nucleus Collisions}

\vspace{1cm}

Marek Ga\'zdzicki\footnote{E--mail: marek@ikf.uni--frankfurt.de}\\
Institut f\"ur Kernphysik, Universit\"at Frankfurt \\
August--Euler--Strasse 6, D - 60486 Frankfurt, Germany\\[0.8cm]

%---- abstract
\begin{minipage}{14cm}
\baselineskip=12pt
\parindent=0.5cm
{\small 

A method to study event--by--event fluctuations
of the "chemical" (particle type) composition of the final
state of high energy collisions is proposed. 
}

\end{minipage}

\end{center}

\begin{center}
{\it }
\end{center}

\vfill
\today
\newpage

\section{Introduction}

Recent data on hadron production in central nucleus--nucleus (A+A) collisions
at the CERN SPS are compatible with 
the hypothesis that a Quark Gluon Plasma \cite{qgp}
is created in the early stage of the interaction \cite{Ga:95}.

This interpretation however requires the assumption that 
the produced matter is close to  thermal and chemical
equilibrium. 
Thus it is important to measure the level of  equilibration
reached in nuclear collisions.

The method to study  
event--by--event fluctuations of 
kinematical variables
("thermal" fluctuations, TF--method)
was proposed
in 1992 \cite{Ga:92} 
and recently used for the analysis of central
Pb+Pb collisions at 158 A$\cdot$GeV by the NA49 Collaboration \cite{Ro:97}.
First results indicate that the analysis of event--by--event 
fluctuations 
adds crucial information concerning the dynamics of A+A collisions
and in particular allows to reject purely  non--equilibrium approches,
like initial state scattering models
\cite{Ga:97}.
The amount of "thermal" fluctuations was recently calculated for
matter in equilibrium \cite{Mr:98}.

In this paper we propose a method to study event--by--event fluctuations
of the particle composition of the final state
of high energy collisions ("chemical" fluctuations, CF--method).
We expect that this method will  allow to determine whether or not
chemical equilibrium is reached in high energy collisions.
As it is based on the TF--method, we start (Section 1) 
from a brief description of
the basic idea and the formalism of the TF--method \cite{Ga:92}.
In Section 3 we introduce the CF--method.
A simple numerical example is presented in Section 4.

\section{Description of the TF--Method}

Let us  suppose  that A+A collisions can be modeled
as a sum of independent nucleon--nucleon (N+N)
interactions.
In this case
event--by-event fluctuations in A+A collisions 
are given by a superposition
of the fluctuations present in  N+N interactions.   
Additional  fluctuations are introduced 
when the number of superimposed N+N interactions varies from
event to event due to e.g. variation in the collision geometry.

A statistical method which allows to remove the influence of trivial geometrical
fluctuations and the effect of averaging over many particle sources was  
introduced in Ref. \cite{Ga:92}.
It was proposed to  quantify the fluctuations
by the so--called $\Phi$ variable.
As an example we consider here a construction 
of  $\Phi$ for the case of transverse momentum
fluctuations.

For every particle $i$ one defines:
\begin{eqnarray*}
z_i = p_{T_i} - \overline{p_T},
\end{eqnarray*}
where $\overline{p_T}$ is the mean transverse momentum calculated for 
all particles from all events (the inclusive mean).
Using $z_i$ one  calculates for every
event
\begin{eqnarray*}
Z = \sum_{i=1}^N z_i,
\end{eqnarray*}
where $N$ is the number of analyzed particles in the event. 
The fluctuation measure\footnote{
In the original paper \cite{Ga:92} the $\Phi_{p_T}$ is called $\Delta D$.
Here we follow notation introduced by the NA49 Collaboration, which
relates the name  of the fluctuation measure
to the variable in which fluctuations are studied.},
$\Phi_{p_T}$,
is then defined as:
\begin{eqnarray}
\Phi_{p_T}  = \sqrt{\frac{\langle Z^2 \rangle}{\langle N \rangle}} -
\sqrt{\; \overline{z^2}},
\end{eqnarray}
where $\langle N \rangle$ and $\langle Z^2 \rangle$ are averages 
(of event--by--event observables) over all
events
and the second term in the r.h.s. is the square root of the second moment
of the inclusive $z$ distribution.

By construction \cite{Ga:92}, the $\Phi_{p_T}$ value
for A+A collisions is equal to the $\Phi_{p_T}$ value
for N+N interactions in the case in which A+A collisions
are pictured as a  sum of independent N+N interactions.
If the  particles are produced independently 
the value of $\Phi_{p_T}$ is equal to zero.

\section{Formulation of the CF--Method}

The TF--method can be  converted into a method allowing to study
event--by--event fluctuations of the "chemical" 
composition (relative number of different hadronic states) 
of the
collisions (CF--method).
In the latter method the basic quantity in which fluctuations are
analyzed is the number of particles of a given type produced in a
single collision. 
This number substitutes  the transverse momentum of 
particles used in the TF--method (see Section 2).

A formal trick which allows for a direct conversion of the TF--method
to the CF--method is based on the substitution of the  kinematical variable 
used for "thermal" fluctuation studies (e.g. $p_T$ as
discussed in Sect. 2) by a  function defined as:
\begin{eqnarray}
 \delta(h^i,h_0) = \left\{ \begin{array} {r@{\quad:\quad}l}
1 & h^i = h_0 \\
0 & h^i \neq h_0 \\
\end{array} \right. 
\end{eqnarray}
where $i$ is a particle index, $h^i$ is the  particle type of the
particle $i$  
and 
$h_0$ is the particle type selected for the fluctuation analysis.
Then the variable $z$ takes the form: 
\begin{eqnarray*}
z_i = \delta(h^i,h_0) - \overline{\delta(h,h_0)},
\end{eqnarray*}
where the  $\overline{\delta(h,h_0)}$ is the mean value of
$\delta(h^i,h_0)$ calculated for all particles from all events 
(the inclusive mean).
$\overline{\delta(h,h_0)}$ 
gives the  probability that any particle of the event ensamble
is of the type
$h_0$.
For every event the sum:
\begin{eqnarray*}
Z = \sum_{i=1}^N z_i,
\end{eqnarray*}
is calculated where the summation runs over all $N$ particles.
Finally the "chemical" fluctuation measure,
$\Phi(h_0)$, can be calculated as:
\begin{eqnarray}
\Phi(h_0) = \sqrt{\frac{\langle Z^2 \rangle}{\langle N \rangle}} -
\sqrt{\; \overline{z^2}},
\end{eqnarray}
where $\langle N \rangle$ and $\langle Z^2 \rangle$ are averages 
(of event--by--event observables) over all
events
and the second term in the r.h.s. is the square root of the second moment
of the inclusive $z$ distribution.

As follows from the construction the important statistical features of the
$\Phi(h_0)$ variable are identical to the features of the $\Phi$ variable
used in the TF--method (e.g. $\Phi_{p_T}$).
We list them below.
\begin{itemize}
\item
For a system which is an independent sum of elementary processes
the value of $\Phi(h_0)$ is 
equal to the value of $\Phi(h_0)$ calculated
for a single elementary process independent of the number of superimposed
elementary processes and its distribution in the analyzed event sample.
\item
In the model in which particles are produced independently  from each
other the value of $\Phi(h_0)$ is equal to zero.
\end{itemize}

\section{Numerical Example}

We first note that
particle production in elementary processes (e.g. p+p interactions)
is correlated not only in  momentum space
\cite{Go:84} but also when the "chemical"
composition is considered.
In order to illustrate this statement by experimental data 
\cite{Ja:75, Bi:92} we show in Fig. 1 the ratio
of the mean $K^0_S$ multiplicity to the multiplicity of negatively charged
hadrons, $n^-$, as a function of $n^-$ for p+p interactions at 200 GeV/c.
The data show 
that the $K^0_S$ multiplicity 
decreases significantly  with the event multiplicity.
A similar correlation can be expected for $K^-$ multiplicity,
and in fact, it is observed in string models of p+p interactions
\cite{Pi:92}. 
It means that the probability that a negatively charged  hadron is  $K^-$ meson
depends on the multiplicity of negatively charged hadrons in the event. 
Thus particles are not produced independently and therefore 
we expect that the value of $\Phi(K^-)$ calculated for negatively
charged hadrons for p+p interactions is not equal to zero.
In order to make  numerical estimation of the effect
we use a  simple parametrization of hadron production in 
p+p interactions at 200 GeV/c.

It is assumed that a dependence of $\langle K^- \rangle / n_- $
on $n_-$ is similar to the dependence of $\langle K^0_S \rangle / n_- $
on $n_-$ shown in Fig. 1.
The multiplicity distribution of negatively charged hadrons is
calculated using the parametrization from Ref. \cite{Ga:91}. 
Further we assume that the multiplicity distribution of $K^-$ mesons
for a fixed multiplicity $n_-$ is given by the binominal distribution
i.e.:
\begin{eqnarray}
P(n_K; n_-) =  { n_- \choose n_K } 
P_K(n_-)^{n_K} (1-P_K(n_-))^{n_- - n_K},
\end{eqnarray}
where $n_K$ is the kaon multiplicity and $P_K(n_-)$
is the probability that a negatively charged hadron is
$K^-$--meson.
Following the data presented in Fig. 1 the $P_K(n_-)$
is parametrized as:        
\begin{eqnarray}
P_K(n_-) = \left\{ \begin{array} {r@{\quad:\quad}l}
P_K(1) - (n_- - 1) \cdot \Delta  & P_K(1) > (n_- - 1) \cdot \Delta  \\
0                          & P_K(1) \le (n_- - 1) \cdot \Delta \\
\end{array} \right.
\end{eqnarray}
In order to study the  dependence of $\Phi(K^-)$ on the correlation
between $P_K(n_-)$ and $n_-$ the parameter $\Delta$ is varied
between 0 (no correlation) and 0.02 (a correlation as suggested
by the data presented
in Fig.~1).
The  parametrizations of $P_K(n_-)$ obtained for these two extreme
values of $\Delta$ 
are indicated by dashed lines in Fig. 1 for $P_K(1) = 0.13$.

The dependence of $\Phi(K^-)$ on the parameter $\Delta$
is shown in Fig. 2 for two values of the parameter $P_K(1)$: 
0.13 (solid line) and 0.26 (dashed line).
For $\Delta = 0$ the value of $\Phi(K^-)$ is equal to zero,
as expected from the definition of $\Phi$.
The value of $\Phi(K^-)$ increases with increasing $\Delta$,
i.e. with the increasing correlation between $K^-$ yield and
the event multiplicity.

Using the above model  ($P_K(1)$ = 0.13 and
$\Delta = 0.02$)
we estimated that the number of events needed to
obtain 10\% statistical error on $\Phi(K^-)$ 
is about $10^5$.

\section{Summary}

We proposed a method to study event--by--event fluctuations
of the "chemical" composition of the final state of high energy 
collisions.
A simple numerical example of fluctuations of the number of 
$K^-$--mesons in the sample of negatively charged hadrons
was considerd.
The method can be used to study a change of the magnitude
of the "chemical" fluctuations when changing the size of the
colliding systems (p+p, p+A and A+A) and/or when changing the
collision energy.

We expect that the method will allow to determine whether or not
chemical equilibration
is reached in high energy nucleus--nucleus collisions. 

\vspace{1cm}

{\bf Acknowledgements}.
I would like to thank Stanis{\l}aw Mr\'owczy\'nski, Reinhard Stock
and Herbert Str\"obele
for discussion and comments.

%\newpage

\begin{figure}[p]
\epsfig{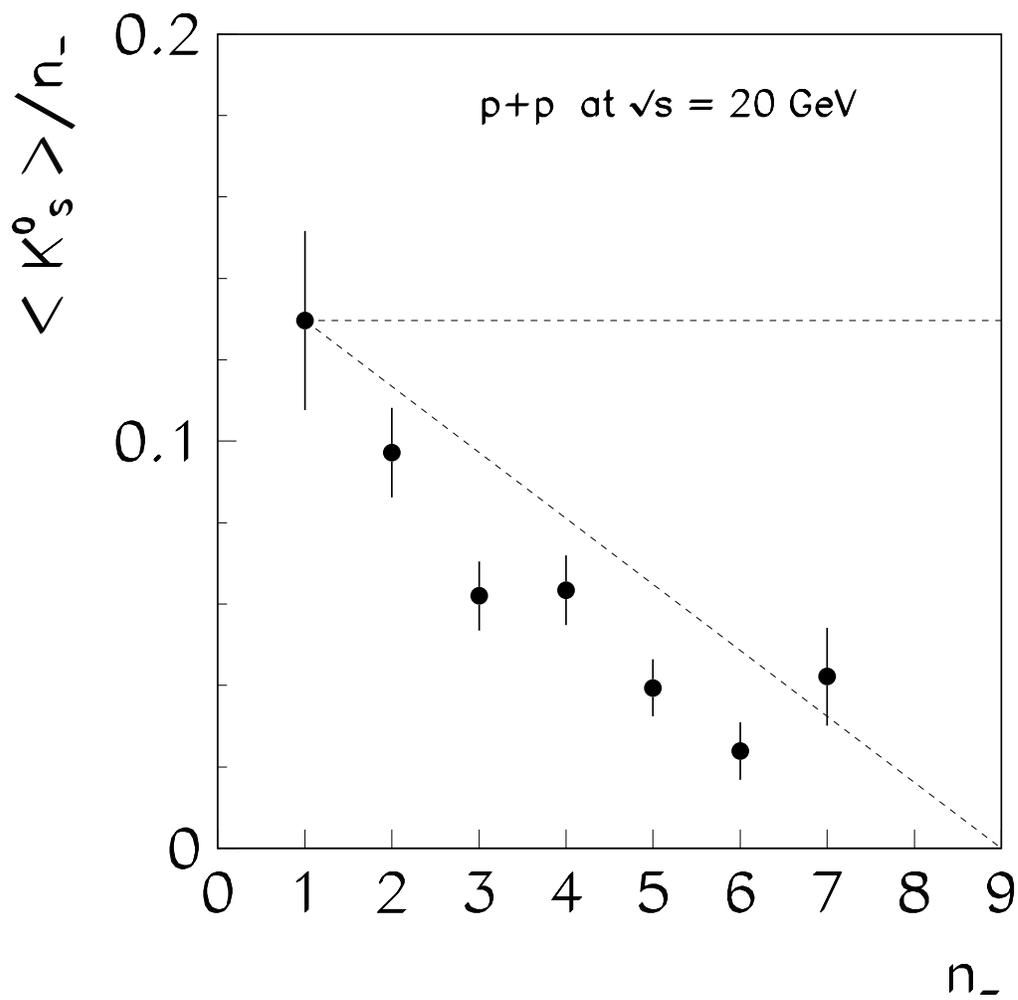}
\caption{ The dependence of the ratio $\langle K^0_S \rangle/n_-$
on negative hadron multiplicity, $n_-$, for p+p interactions at
$\sqrt{s}$ = 20 GeV.
}
\label{fig1}
\end{figure}

\begin{figure}[p]
\epsfig{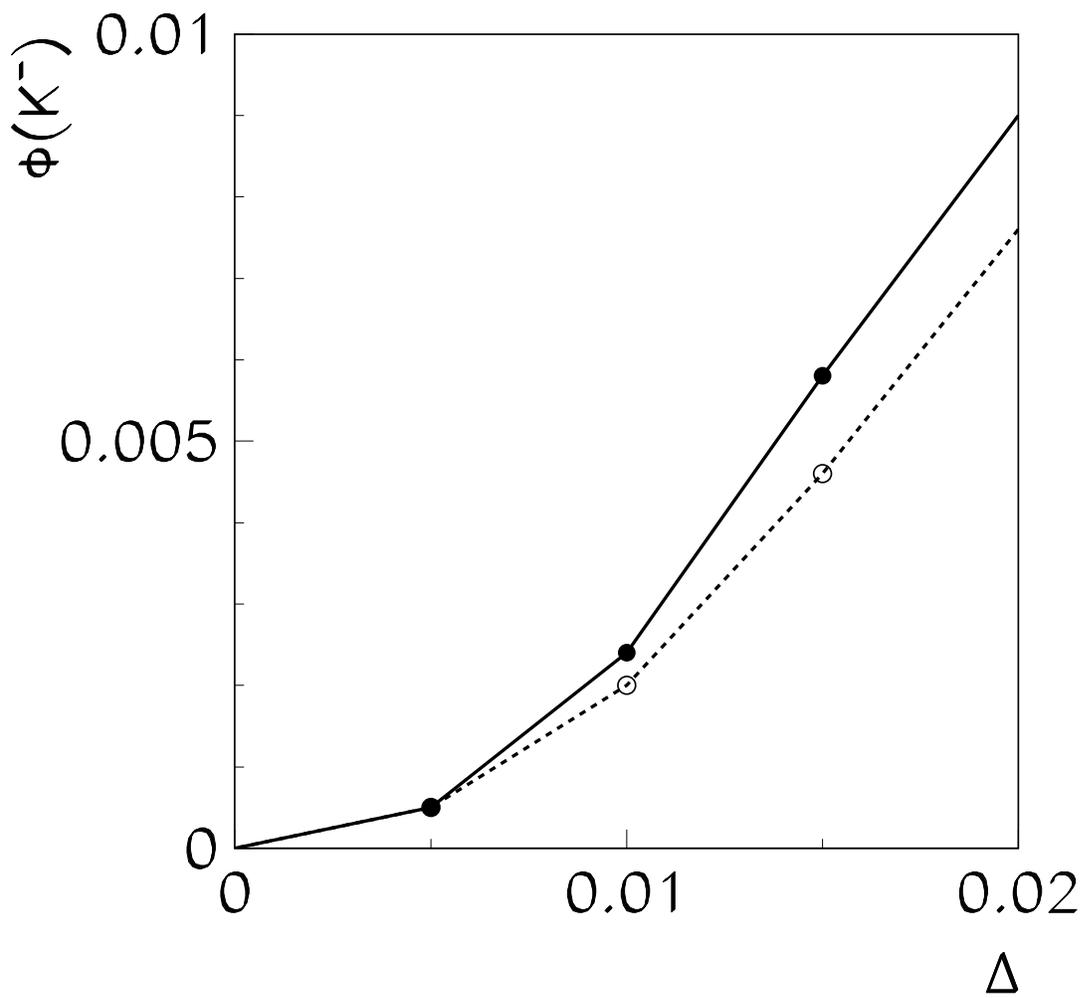}
\caption{ The dependence of the  fluctuation measure 
$\Phi(K^-)$ 
on the paremeter $\Delta$ (see text) for two values of the parameter $P_K(1)$:
0.13 (solid line) and 0.26 (dashed line).
}
\label{fig2}
\end{figure}


\begin{thebibliography}{99}


\bibitem{qgp} J. C. Collins and M. J. Perry, Phys. Rev. Lett.
{\bf 34} (1975) 151, \\
E. V. Shuryak, Phys. Rep. {\bf C61} (1980)
71 and {\bf C115} (1984) 151.
995.

\bibitem{Ga:95} M. Ga\'zdzicki, Z. Phys. {\bf C66} (1995) 659, \\
M. Ga\'zdzicki and D. R\"ohrich, Z. Phys. {\bf C71} (1996) 55, \\
M. Ga\'zdzicki,   
J. Phys. {\bf G23} (1997) 1881.

\bibitem{Ga:92} M.\ Ga\'zdzicki, St.\ Mr\'owczy\'nski, Z.\ Phys.\ {\bf C54} (1992) 127.

\bibitem{Ro:97} G.\ Roland et al. (NA49 Collab.), Proceedings of the Hirschegg Workshop on QCD Phase
Transitions, 1997 page 309.

\bibitem{Ga:97}
M. Ga\'zdzicki, A. Leonidov, G. Roland, 
Franfurt University Report, IKF--HENPG/5--97 and
hep--ph/9711422, to be published in Eur. Phys. J. {\bf C}.

\bibitem{Mr:98}
St. Mr\'owczy\'nski, {\it Transverse momentum and energy correlations
in the equilibrium system from high energy collisions}, nucl--th/9806089,
to be published in Phys. Lett {\bf B}.



\bibitem{Go:84} 
A. I. Golokhvastov, Z. Phys. {\bf C26} (1984) 469.

\bibitem{Ja:75}
K. Jaeger et al., Phys. Rev. {\bf D9} (1975) 2405.

\bibitem{Bi:92}
H. Bia{\l}kowska, M. Ga\'zdzicki, W. Retyk, E. Skrzypczak,
Z. Phys. {\bf C55} (1992) 491.

\bibitem{Pi:92}
H. Pi, Computer Physics Commun. {\bf 71} (1992) 173.

\bibitem{Ga:91}
M. Ga\'zdzicki, R. Szwed, G. Wrochna, A. K. Wr\'oblewski,
Mod. Phys. Lett. {\bf A6} (1991) 981.


\end{thebibliography}
\end{document}